\documentclass[a4paper,english]{revtex4}
\usepackage[T1]{fontenc}
\usepackage[latin9]{inputenc}
\usepackage{graphicx}
\usepackage{amssymb}
\usepackage{esint}

\makeatletter

\usepackage{babel}
\makeatother

\begin{document}

\title{Density oscillations in multi-component molecular mixtures}

\author{M. Apostol}

\affiliation{Department of Theoretical Physics, Institute for Atomic Physics,
Magurele-Bucharest}

\address{Magurele-Bucharestr MG-6, POBox MG-35, Romania}

\email{apoma@theory.nipne.ro}

\begin{abstract}
The excitation spectrum of the density collective oscillations is
computed for multi-component molecular mixtures both with Coulomb
and (repulsive) short-range interactions. Distinct sound-like excitations
appear, governed by the short-range interaction, which differ from
the ordinary hydrodynamic sound. The dielectric function and the structure
factor are also calculated. The \char`\"{}two-sounds phenomenon\char`\"{}
can be understood by means of the predictions of this model. 
\end{abstract}



\maketitle
This paper is motivated by the \char`\"{}two-sounds anomaly\char`\"{}
persistently reported over the years in water, either in normal conditions
or undercooled,\citet*{key-1}-\citet{key-6} as well as in other
liquid molecular mixtures. Inelastic neutron, $X$-ray, Brillouin
and, more recently, ultraviolet scattering, either in ordinary or
in heavy water, seem to indicate an additional, faster, higher-frequency
sound, propagating with velocity $\simeq3000m/s$ up to intermediate
wavevectors (mean inter-molecular distance in water is $\simeq3\textrm{\AA}$),
beside the ordinary hydrodynamic sound propagating with velocity $\simeq1500m/s$.
A dispersionless mode ($\simeq10^{13}s^{-1}$) was also reported sometimes\citep{key-3},\citep{key-5}
(as well as no additional sound\citep{key-7}). The phenomenon is
also documented both by simulations of molecular dynamics and experimental
data in binary mixtures with large mass difference (metallic alloys,
rare-gas mixtures).\citep{key-8}-\citep{key-17}

We show herein that such a \char`\"{}two-sounds anomaly\char`\"{}
may appear in interacting molecular systems with (repulsive) short-range
interaction. Such a model could reasonably be related to liquid water
(or other physical systems as those indicated above). The velocity
of the sound-like excitations is independent of temperature, in contrast
with the velocity of the hydrodynamic sound which is governed by the
adiabatic compressibility, and thus temperature-dependent. In addition,
the plasma-like branch of the spectrum due to the Coulomb interaction
may appear as another sound-like mode for shorter wavelengths and
weak Coulomb coupling. We report here also the computation of the
dielectric function and the structure factor within such a model. 

We start with the well-known representation of the particle density
\begin{equation}
n(\mathbf{r})=\frac{1}{V}\sum_{i}\delta(\mathbf{r}-\mathbf{r}_{i})=\frac{1}{V}\sum_{\mathbf{q}}e^{i\mathbf{qr}}\sum_{i}e^{-i\mathbf{q}\mathbf{r}_{i}}\,\,\,\label{1}\end{equation}
 for a collection of $N$ particles enclosed in volume $V$, where
$\mathbf{r}_{i}$ denotes the position of the $i$-th particle. We
consider a small displacement $\mathbf{r}_{i}\rightarrow\mathbf{r}_{i}+\mathbf{u}(\mathbf{r}_{i})$
in these positions, as given by a displacement field $\mathbf{u}(\mathbf{r}_{i})$,
such that the particle density becomes \begin{equation}
\widetilde{n}(\mathbf{r})=\frac{1}{V}\sum_{\mathbf{q}}e^{i\mathbf{qr}}\sum_{i}e^{-i\mathbf{q}\left[\mathbf{r}_{i}+\mathbf{u}(\mathbf{r}_{i})\right]}=\frac{1}{V}\sum_{\mathbf{q}}e^{i\mathbf{qr}}\sum_{i}e^{-i\mathbf{q}\mathbf{r}_{i}}\left[1-i\mathbf{qu}(\mathbf{r}_{i})+...\right]\,\,\,\label{2}\end{equation}
 for $\mathbf{qu}(\mathbf{r}_{i})\ll1$. Now we employ a Fourier representation
\begin{equation}
\mathbf{u}(\mathbf{r}_{i})=\frac{1}{\sqrt{N}}\sum_{\mathbf{q}}\mathbf{u}(\mathbf{q})e^{i\mathbf{q}\mathbf{r}_{i}}\,\,\,\label{3}\end{equation}
 as well as the well-known random-phase approximation \begin{equation}
\sum_{i}e^{i(\mathbf{q}-\mathbf{q}')\mathbf{r_{i}}}=N\delta_{\mathbf{q},\mathbf{q}'}\,\,\,\label{4}\end{equation}
 to get \begin{equation}
\widetilde{n}(\mathbf{r})=n-in\frac{1}{\sqrt{N}}\sum_{\mathbf{q}}e^{i\mathbf{qr}}\mathbf{qu}(\mathbf{q})\,\,\,,\label{5}\end{equation}
 where $n=N/V$ is the particle density. By comparing equations (\ref{1})
and (\ref{5}), we can see that the small change in the density can
be represented as \begin{equation}
\widetilde{n}(\mathbf{r})-n=\delta n(\mathbf{r})=-ndiv\mathbf{u}(\mathbf{r})\,\,\,,\label{6}\end{equation}
and its Fourier transform $\delta n(\mathbf{q})=-in\mathbf{q}\mathbf{u}(\mathbf{q})$. 

We apply this displacement-field approach to a multi-component molecular
mixture consisting of several species labelled by $i$, each with
$N_{i}$ particles in volume $V$, mass $m_{i}$ and electric charge
$ez_{i}$, where $-e$ is the electron charge and $z_{i}$ is a reduced
effective charge, interacting through Coulomb potentials $\varphi_{ij}$
and short range potentials $\chi_{ij}$. The mixture is subjected
to the neutrality condition $\sum_{i}n_{i}z_{i}=0$, where $n_{i}=N_{i}/V$
is the particle density of the $i$-th species. We consider elementary
excitations of the particle density, whose interaction energy is given
by \begin{equation}
U=\frac{1}{2}\sum_{ij}\int d\mathbf{r}d\mathbf{r}'\left[\varphi_{ij}(\mathbf{r}-\mathbf{r}')+\chi_{ij}(\mathbf{r}-\mathbf{r}')\right]\delta n_{i}(\mathbf{r})\delta n_{j}(\mathbf{r}')\,\,\,,\label{7}\end{equation}
where $\varphi_{ij}=e^{2}z_{i}z_{j}/\left|\mathbf{r}-\mathbf{r}'\right|$
and $\delta n_{i}(\mathbf{r})$ denotes a small density disturbance
which preserves the neutrality. According to equation (\ref{6}) it
can be represented as $\delta n_{i}=-n_{i}div\mathbf{u}_{i}$, where
$\mathbf{u}_{i}$ is the displacement field. We use the Fourier transforms
\begin{equation}
\delta n_{i}(\mathbf{r})=\frac{1}{\sqrt{N}}\sum_{\mathbf{q}}\delta n_{i}(\mathbf{q})e^{i\mathbf{qr}}\,\,,\,\,\varphi(\mathbf{r})=\frac{1}{V}\sum_{\mathbf{q}}\varphi(\mathbf{q})e^{i\mathbf{qr}}\,\,\,,\label{8}\end{equation}
where $N=\sum_{i}N_{i}$ is the total number of particles, $\varphi(\mathbf{r})=e^{2}/r$
and $\varphi(\mathbf{q})=\varphi(q)=4\pi e^{2}/q^{2}$. A similar
Fourier transform is employed for the displacement field $\mathbf{u}_{i}$,
which leads to $\delta n_{i}(\mathbf{q})=-in_{i}\mathbf{q}\mathbf{u}_{i}(\mathbf{q})$.
We can see that only the longitudinal components $u_{i}(\mathbf{q})$
of the displacement field are relevant, so we may write $\mathbf{u}_{i}(\mathbf{q})=(\mathbf{q}/q)u_{i}(\mathbf{q})$,
$\delta n_{i}(\mathbf{q})=-iqu_{i}(\mathbf{q})$, with $\delta n_{i}^{*}(-\mathbf{q})$=$\delta n_{i}(\mathbf{q})$,
$\mathbf{u}_{i}^{*}(-\mathbf{q})=\mathbf{u}_{i}(\mathbf{q})$ and
$u_{i}^{*}(-\mathbf{q})=-u_{i}(\mathbf{q})$. Making use of the Fourier
transforms introduced above, the interaction $U$ given by equation
(\ref{7}) can be written as \begin{equation}
U=-\frac{1}{2n}\sum_{ij\mathbf{q}}n_{i}n_{j}q^{2}\left[\varphi_{ij}(q)+\chi_{ij}(q)\right]u_{i}(\mathbf{q})u_{j}(-\mathbf{q})\,\,\,,\label{9}\end{equation}
 where $\varphi_{ij}(q)=z_{i}z_{j}\varphi(q)$ and $n=N/V$ is the
total density of particles. We assume a weak $q$-dependence of $\chi_{ij}(q)$,
as for short-range potentials.

Similarly, the kinetic energy associated with the coordinates $u_{i}$
is given by \begin{equation}
T=-\frac{1}{2n}\sum_{i\mathbf{q}}m_{i}n_{i}\dot{u}_{i}(\mathbf{q})\dot{u}_{i}(-\mathbf{q})\,\,.\label{10}\end{equation}
 In addition, we introduce an external field $\phi(\mathbf{r})$,
coupled to the electrical charges, which gives rise to the interaction
\begin{equation}
V=-i\frac{e}{n}\sum_{i\mathbf{q}}n_{i}z_{i}q\phi(\mathbf{q})u_{i}(-\mathbf{q})\,\,.\label{11}\end{equation}

The equations of motion corresponding to the lagrangian $L=T-U-V$
are given by \begin{equation}
m_{i}\ddot{u}_{i}+4\pi e^{2}z_{i}\sum_{j}z_{j}n_{j}u_{j}+q^{2}\sum_{j}\chi_{ij}n_{j}u_{j}=-iqez_{i}\phi\,\,\,,\label{12}\end{equation}
 where we dropped out the argument $\mathbf{q}$ in $u_{i}(\mathbf{q})$
and $\phi(\mathbf{q})$ and neglect the weak $q$-dependence of $\chi_{ij}(q)=\chi_{ij}$.
In order to simplify these equations we take the same (repulsive)
short-range potentials for all species, $\chi_{ij}=\chi>0$, and analyze
first the homogeneous system of equations given by (\ref{12}). We
introduce the notations $a=4\pi e^{2}$, $b=q^{2}\chi$, \begin{equation}
S_{1}=\sum_{i}\frac{z_{i}^{2}n_{i}}{m_{i}}\,\,,\,\, S_{2}=\sum_{i}\frac{n_{i}}{m_{i}}\,\,,\,\, S_{3}=\sum_{i}\frac{z_{i}n_{i}}{m_{i}}\,\,\,,\label{13}\end{equation}
and \begin{equation}
x=\frac{1}{n}\sum_{i}z_{i}n_{i}u_{i}\,\,,\,\, y=\frac{1}{n}\sum_{i}n_{i}u_{i}\,\,.\label{14}\end{equation}
 Making use of these notations, the homogeneous system of equations
(\ref{12}) can be written as \begin{equation}
\begin{array}{c}
\left(-\omega^{2}+aS_{1}\right)x+bS_{3}y=0\,\,\,,\\
\\aS_{3}x+\left(-\omega^{2}+bS_{2}\right)y=0\,\,.\end{array}\label{15}\end{equation}
 In addition, we have \begin{equation}
\omega^{2}u_{i}=\frac{anz_{i}}{m_{i}}x+\frac{bn}{m_{i}}y\,\,.\label{16}\end{equation}

The spectrum of frequencies $\omega$ of the system of equations (\ref{15})
can be obtained straightforwardly. It is given by \begin{equation}
\omega_{1,2}^{2}=\frac{1}{2}\left[aS_{1}+bS_{2}\pm\sqrt{a^{2}S_{1}^{2}+2ab\left(2S_{3}^{2}-S_{1}S_{2}\right)+b^{2}S_{2}^{2}}\right]\,\,.\label{17}\end{equation}

The $\omega_{1}$-branch in equation (\ref{17}) (corresponding to
the plus sign) represents the plasmonic excitations. In the long wavelength
limit it reads \begin{equation}
\omega_{1}^{2}=aS_{1}+bS_{3}^{2}/S_{1}=\omega_{p}^{2}+bS_{3}^{2}/S_{1}\,\,\,,\,\,\, q\rightarrow0\,\,\,,\label{18}\end{equation}
where $\omega_{p}$, given by \begin{equation}
\omega_{p}^{2}=aS_{1}=4\pi e^{2}\sum_{i}\frac{z_{i}^{2}n_{i}}{m_{i}}\,\,\,,\label{19}\end{equation}
 is the plasma frequency. For shorter wavelengths the $\omega_{1}$-branch
approaches an asymptote given by \begin{equation}
\omega_{1}^{2}\simeq bS_{2}+aS_{3}^{2}/S_{2}\,\,\,,\,\,\, q\rightarrow\infty.\label{20}\end{equation}

The $\omega_{2}$-branch in equation (\ref{17}) (corresponding to
the minus sign) represents sound-like excitations. In the long wavelength
limit it is given by \begin{equation}
\omega_{2}^{2}=\left(S_{2}-S_{3}^{2}/S_{1}\right)b=v_{s}^{2}q^{2}\,\,\,,\,\,\, q\rightarrow0\,\,\,,\label{21}\end{equation}
 where \begin{equation}
v_{s}=\sqrt{\left(S_{2}-S_{3}^{2}/S_{1}\right)\chi}\,\,\,\label{22}\end{equation}
 is the corresponding sound velocity. We can see easily, by applying
the Schwarz-Cauchy inequality to the vectors $a_{i}=\sqrt{n_{i}/m_{i}}$
and $b_{i}=z_{i}\sqrt{n_{i}/m_{i}}$, that $v_{s}^{2}$ is always
positive ($(S_{2}-S_{3}^{2}/S_{1})\geq0$). For shorter wavelengths
the $\omega_{2}$-branch of the spectrum approaches an horizontal
asymptote given by \begin{equation}
\omega_{2}^{2}\simeq\left(1-S_{3}^{2}/S_{1}S_{2}\right)\omega_{p}^{2}\,\,\,,\,\,\, q\rightarrow\infty\,\,.\label{23}\end{equation}
In the limit of vanishing Coulomb coupling ($a\rightarrow0$) the
sound-branch of the spectrum becomes $\omega_{2}^{2}=bS_{2}=v_{s}^{2}q^{2}$,
where \begin{equation}
v_{s}^{2}=\chi S_{2}=\chi\sum_{i}(n_{i}/m_{i})\,\,\,,\label{24}\end{equation}
an expression which holds also for the same mass $m_{i}=m$ for all
particles (one component), due to the neutrality condition ($S_{3}=0$). 

The above elementary excitations, which are governed by interaction,
are non-equilibrium collective modes which might be termed density
\char`\"{}kinetic\char`\"{} modes.\citep{key-18} The sound-like excitations
($\omega_{2}$-branch in equation (\ref{17})) may be called \char`\"{}densitons\char`\"{},
in order to distinguish them from plasmons ($\omega_{1}$-branch in
equation (\ref{17})) and from the ordinary sound. They may correspond
to the density collective modes suggested by Zwanzig for classical
liquids.\citep{key-19} We emphasize that these sound-like excitations
are distinct from the ordinary hydrodynamic sound. 

Indeed, the interaction corresponding to the latter can be written
as \begin{equation}
U=\frac{1}{2\kappa}\int d\mathbf{r}\left[div\mathbf{u}(\mathbf{r})\right]^{2}=-\frac{1}{2\kappa n}\sum_{\mathbf{q}}q^{2}u(\mathbf{q})u(-\mathbf{q})\,\,\,,\label{25}\end{equation}
where $\kappa=-(1/V)(\partial V/\partial p)_{S}$ is the adiabatic
compressibility ($p$ denotes the pressure and $S$ stands for entropy).
The above equation is derived by making use of the change $\delta V=-V(\delta n/n)=Vdiv\mathbf{u}$
in volume. We emphasize that for thermodynamic equilibrium we have
only one displacement field $\mathbf{u}(\mathbf{r})$. Equation (\ref{25})
together with the kinetic energy given by equation (\ref{10}) for
$u_{i}(\mathbf{q})=u(\mathbf{q})$ leads to the sound branch $\omega_{0}^{2}=v_{0}^{2}q^{2}$,
corresponding to the ordinary sound propagating with a velocity $v_{0}$
given by \begin{equation}
v_{0}^{2}=\left(\kappa\sum_{i}n_{i}m_{i}\right)^{-1}\,\,.\label{26}\end{equation}
For $m_{i}=m$ (one component) the above equation gives the well-known
velocity $v_{0}=1/\kappa nm$ of the ordinary sound. As it is well-known,
it has a slight temperature dependence, through the compressibility,
in contrast with the velocity $v_{s}$ given above for the sound-like
excitations. For an electrically neutral multi-component mixture it
can be shown easily that $v_{s}^{2}/v_{0}^{2}\geq n^{2}\chi\kappa$.

If we apply equations (\ref{24}) and $(\ref{26})$ to both ordinary
and heavy water (one component, neutral molecule), and assume that
interaction $\chi$ and, respectively, the compressibility $\kappa$
are the same for the two kinds of water, we can see that the two sound
velocities $v_{s}$ and $v_{0}$ exhibit a slight isotopic effect,
while their ratio $v_{s}/v_{0}=n\sqrt{\chi\kappa}$ does not exhibit
such an isotopic effect, in agreement with experimental data. In this
case we may take $v_{0}=1500m/s$ and $v_{s}=3000m/s$ from experimental
data and get the interaction parameter $\chi\simeq60eV\cdot\textrm{\AA}^{3}$
(for a mean inter-molecular spacing $\simeq3\textrm{\AA}$) . A similar
picture, given by equations (\ref{24}) and (\ref{26}), may apply
to rare-gas mixtures, while for metallic alloys the Coulomb coupling
must be taken into account (and equation (\ref{22}) employed). 

If we assume the existence of a dispersionless mode in water, then
we may consider that water molecule is dissociated to some extent,
and its components have an electric charge, such that the plasmonic
mode given by equation (\ref{19}) can be identified with such a dispersionless
mode. Various models of dissociation of the water molecule are known,
like $OH^{-}-H^{+}$ or $OH^{-}-H_{3}O^{+}$. In all cases a certain
mobility of the $H^{+}$ (hydrogen) cations and $O^{-}$ (oxygen)
anions is implied. We assume here that the dynamics of liquid water
has a plasma-like component consisting of $H^{+z}$ cations with density
$2n$ and mass $m$ (proton mass) and $O^{-2z}$ anions with density
$n$ and mass $M=16m$, where $n$ is the density of water. The excitation
spectrum given by equations (\ref{17}) for such an $O^{-2z}-H^{+z}$
plasma is shown in Fig. 1. %
\begin{figure}
\noindent \begin{centering}
\includegraphics[clip,scale=0.4]{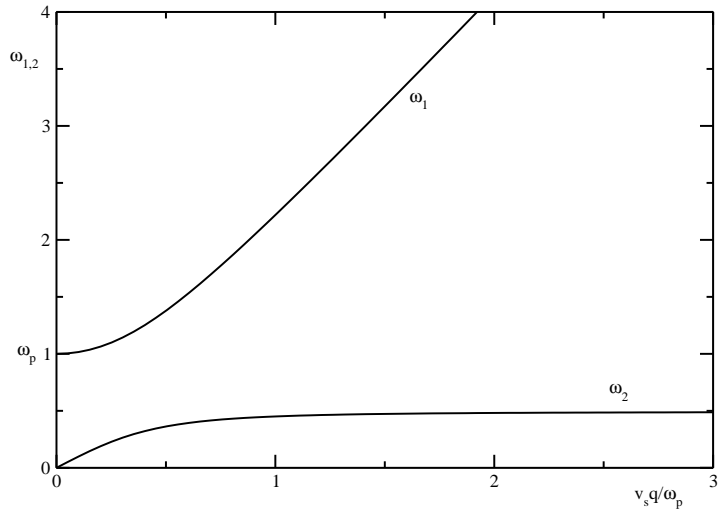}
\par\end{centering}

\caption{Spectrum of density excitations given by equation (\ref{17}) for
the $O^{-2z}-H^{+z}$ plasma.}

\end{figure}
 Taking $\omega=10^{13}s^{-1}$($\simeq5meV$) of the dispersionless
mode\citep{key-3},\citep{key-5},\citep{key-7} as the plasma frequency
$\omega_{p}$ given by $\omega_{p}^{2}=16\pi ne^{2}z^{2}/\mu$ (equation
(\ref{19})), where $\mu=2mM/(M+2m)$ is the reduced mass, we get
$z\simeq3\times10^{-2}$. The velocity of the hydrodynamic sound is
given by $v_{0}=1/\sqrt{\kappa n(M+2m)}$ according to equation (\ref{26})
and the velocity of the sound-like excitations is given by $v_{s}=\sqrt{9n\chi/(M+2m)}$
from equation (\ref{22}). We can see that both velocities exhibit
an isotopic effect, but their ratio $v_{s}/v_{0}=3n\sqrt{\chi\kappa}$
does not, in agreement with the experimental data. From $v_{s}=3000m/s$
we derive the interaction $\chi\simeq7eV\cdot\textrm{\AA}^{3}$. Similar
results are obtained for other forms of dissociation, like $OH^{-}-H^{+}$
or $OH^{-}-H_{3}O^{+}$. In this respect, the $O^{-2z}-H^{+z}$ plasma
model can be viewed as an average, effective model for various plasma
components that may exist in water. 

According to equation (\ref{20}), for shorter wavelengths the $\omega_{1}$-branch
approaches an asymptote given by $\omega_{1}^{2}\sim bS_{2}+aS_{3}^{2}/S_{2}$.
In the limit of weak Coulomb coupling this $\omega_{1}$-branch may
appear as an \char`\"{}anomalous\char`\"{} sound given by \begin{equation}
\omega_{a}=\sqrt{bS_{2}}=v_{a}q\,\,\,,\label{27}\end{equation}
propagating with velocity \begin{equation}
v_{a}=\sqrt{S_{2}\chi}=\frac{1}{\sqrt{1-S_{3}^{2}/S_{1}S_{2}}}v_{s}\,\,\,\label{28}\end{equation}
(which is always a positive qauntity). This additional, anomalous
sound is always faster than the sound-like excitations propagating
with velocity $v_{s}$, since \begin{equation}
\frac{v_{a}}{v_{s}}=\frac{1}{\sqrt{1-S_{3}^{2}/S_{1}S_{2}}}>1\,\,.\label{29}\end{equation}

It is worth noting that the molecular dynamics studies which originary
predicted such a fast, anomalous sound\citep{key-20} employed indeed
a Coulomb interaction and a short-range one. We note, however, that
the velocity $v_{a}$ as given by (\ref{28}) does not depend on the
Coulomb coupling. In the plasma model for water discussed above the
ratio $v_{a}/v_{0}$ is approximately $2$ ($\simeq\sqrt{2M/9m+5/9}$),
but it exhibits an isotopic effect, which does not seem to be supported
by the experimental data. 

It is easy to derive the dielectric function in the limit of long
wavelengths from equation (\ref{12}). Indeed, for charged particles
equation $\delta n_{i}=-n_{i}div\mathbf{u}_{i}$ is equivalent with
Maxwell equation $div\mathbf{E}_{i}=4\pi q_{i}\delta n_{i}$, where
the electric field is given by $\mathbf{E}_{i}=-4\pi q_{i}n_{i}\mathbf{u}_{i}$
and $q_{i}=ez_{i}$ is the electric charge of the $i$-th species.
It follows that the internal field is given by \begin{equation}
E_{int}=-4\pi e\sum_{i}z_{i}n_{i}u_{i}\,\,\label{30}\end{equation}
We get easily this field from equations (\ref{12}), \begin{equation}
E_{int}=-iq\phi\frac{\omega_{p}^{2}}{\omega^{2}-\omega_{p}^{2}}\,\,\,\label{31}\end{equation}
in the long wavelength limit (it is proportional to $x$ given by
equation (\ref{14})). The dielectric function is defined by $D=\varepsilon E=\varepsilon(D+E_{int})$,
where $D=-iq\phi$ is the external field (electric displacement).
We get the plasma dielectric function \begin{equation}
\varepsilon=1-\omega_{p}^{2}/\omega^{2}\,\,\,,\label{32}\end{equation}
as expected. It exhibits an absorption edge ($\omega_{p}$) for very
low frequencies. In the static limit it is reasonable to admit the
existence of an additional internal field of intrinsic polarizability
which removes the $\omega=0$ singularity. 

We pass now to the calculation of the structure factor. From equation
(\ref{16}) we can see that the displacement $u_{i}$ is a superposition
of the two eigenvectors of the system of equations (\ref{15}), which
oscillate with eigenfrequencies $\omega_{1,2}$, respectively. It
follows that these coordinates are those of linear harmonic oscillators
with the potential energy of the form $m_{i}\omega^{2}u_{i}^{2}/2$.
The statistical distribution of the coordinates $u_{i}$ in the classical
limit is given by $dw\sim\exp(-m_{i}\omega^{2}u_{i}^{2}/2T)du_{i}$,
where $T$ denotes the temperature. We get the thermal averages \begin{equation}
\left\langle u_{i}u_{j}\right\rangle =\frac{T}{m_{i}\omega^{2}}\delta_{ij}\,\,.\label{33}\end{equation}
 On the other hand the structure factor defined by \begin{equation}
\begin{array}{c}
S(q,\omega)=\frac{1}{2\pi}\int d\mathbf{r}d\mathbf{r}'dt\left\langle \delta n(\mathbf{r},t)\delta n(\mathbf{r}',0)\right\rangle e^{i\mathbf{q}(\mathbf{r}-\mathbf{r}')-i\omega t}=\\
\\=\frac{N}{2\pi n^{2}}\int dt\left\langle \delta n(\mathbf{q},t)\delta n(-\mathbf{q},0)\right\rangle e^{-i\omega t}\,\,\,\end{array}\label{34}\end{equation}
(we leave aside the central peak) can be written as \begin{equation}
S(q,\omega)=\frac{Nq^{2}}{2\pi n^{2}}\int dt\sum_{ij}n_{i}n_{j}\left\langle u_{i}(t)u_{j}(0)\right\rangle e^{-i\omega t}\,\,.\label{35}\end{equation}
Writing \begin{equation}
u_{i}=u_{i}^{(1)}e^{i\omega_{1}t}+u_{i}^{(2)}e^{i\omega_{2}t}\,\,\,\label{36}\end{equation}
 and making use of equation (\ref{33}) we get the structure factor
\begin{equation}
S(q,\omega)=NTq^{2}\left(\sum_{i}n_{i}^{2}/n^{2}m_{i}\right)\left[\frac{1}{\omega_{1}^{2}}\delta(\omega-\omega_{1})+\frac{1}{\omega_{2}^{2}}\delta(\omega-\omega_{2})\right]\,\,.\label{37}\end{equation}
 We can see that the relevant sound contributions read \begin{equation}
S(q,\omega)\simeq\frac{NT}{v_{s,a}^{2}}\left(\sum_{i}n_{i}^{2}/n^{2}m_{i}\right)\delta(\omega-v_{s,a}q)\,\,.\label{38}\end{equation}
The relaxation and damping effects can be included in the above expressions
of the structure factor. As it is well-known, they amount to representing
the $\delta$-functions by lorentzians. 

The short-range interaction $\chi$ can be generalized to an interaction
matrix $\chi_{ij}$ with distinct elements for each pair of species.
In this case, the excitation spectrum of the density oscillations
may exhibit multiple branches in general, for a multi-component mixture.
In addition, it may have special features, like a dip in the plasmonic
branch, or negative velocity for the sound-like excitations, which
may indicate either an anomalous behaviour or unphysical situations,
depending on the mutual magnitudes of the short-range potentials $\chi_{ij}$. 

Now it is worthwhile commenting upon the validity of the approach
presented above. If we keep higher-order terms in the expansion given
by equation (\ref{2}) (\emph{i.e.} for moderate values of $\mathbf{q}\mathbf{u}_{i}$)
then additional interactions appear in equation (\ref{9}), which
leads to finite lifetimes for the density excitations. This means
that for larger wavevectors $\mathbf{q}$ these excitations are not
anymore well-defined excitations, as expected. Making use of equation
(\ref{33}) we can estimate the mean product $qu_{i}$ for the sound-like
branch as $qu_{i}\sim\sqrt{T/m_{i}v_{s}^{2}}$ , where the velocity
$v_{s}$ is given by equation (\ref{22}). This gives rather small
values for $qu_{i}$. For instance, for water we get $qu\sim0.5$
(at room temperature), which shows that the wavevector $q$ may take
reasonable large values providing the displacement $u$ is sufficiently
small. For the plasmonic branch, the condition $qu_{i}\ll1$ gives
a cutoff wavevector $q_{c}^{i}\simeq\sqrt{m_{i}\omega_{p}^{2}/T}$
for large $\omega_{p}$; for small values of the plasma frequency
the condition becomes $qu_{i}\sim\sqrt{T/m_{i}v_{a}^{2}}\ll1$. 

Another source of finite lifetime for the density excitations arises
from the kinetic term. Indeed, under the displacement $\mathbf{r}_{ik}\rightarrow\mathbf{r}_{ik}+\mathbf{u}_{i}(\mathbf{r}_{ik})$,
where $\mathbf{r}_{ik}$ is the position of the $k$-th particle in
the $i$-th species, a mixed term\begin{equation}
H_{int}=\sum_{ik}m_{i}\mathbf{v}_{ik}\dot{\mathbf{u}}_{i}(\mathbf{r}_{ik})\,\,\,\label{39}\end{equation}
appears in the kinetic term, where $\mathbf{v}_{ik}=\dot{\mathbf{r}}_{ik}$
is the velocity of the $ik$-particle. It is easy to get an upper
bound for this term, by using the Schwarz-Cauchy inequality. It is
given by $N\left\langle m_{i}v_{ik}^{2}\right\rangle ^{1/2}\left\langle m_{i}\dot{u}_{i}^{2}\right\rangle ^{1/2}$
or, by making use of (\ref{33}), $\sqrt{\varepsilon T}$ per particle,
where $\varepsilon$ represents the mean kinetic energy (which depends
on temperature, in principle). This estimation can be taken as an
uncertainty in energy, leading to a lifetime $\tau\simeq\hbar/\sqrt{\varepsilon T}$
and a corresponding meanfree path $\Lambda=v_{s}\tau$ for the sound-like
excitations. For wavelengths $\lambda$ much longer than the meanfree
path, \emph{i.e. }for wavevectors $q$ such as  $q\ll1/v_{s}\tau$
we are in the collision-like regime ($\omega_{2}\tau\ll1$), and the
collisions can establish the thermodynamic equilibrium (hydrodynamic
regime). In this case the ordinary sound can be propagated (with velocity
$v_{0}$). For $q\gg1/v_{s}\tau$ we are in the collisionless regime,
the ordinary sound is absorbed, and the non-equilibrium sound-like
excitations (\char`\"{}densitons\char`\"{}) can be propagated (with
velocity $v_{s}$). Unfortunately, it is difficult to have a reliable
estimation of the energy $\varepsilon$, and so of the threshold wavevector
$q_{t}=1/v_{s}\tau=\sqrt{\varepsilon T}/\hbar v_{s}$. For $\varepsilon=10meV$
(and $v_{s}=3000m/s$, $T=300K$) we get $q_{t}\simeq0.1\textrm{\AA}^{-1}$,
which is in a reasonable order-of-magnitude agreement with the experimental
data.\citep{key-1}-\citep{key-6},\citep{key-11},\citep{key-12},\citep{key-15}
It is interesting to note that if we apply this estimation to weakly-interacting
gases, where we may take $\varepsilon\sim T$, we get a high value
of the threshold wavevector $q_{t}\sim T/\hbar v_{s}$, since $v_{s}$
is very small (the short-range interaction is weak). We may say that
in gases there is very unlikely to exist the sound-like excitations;
it is only the ordinary sound that exists. On the contrary, the collision-like
regime is quite unlikely in ordinary solids, so we have there sound-like
excitations and to a much lesser extent ordinary sound.

Finally, we note that the collective excitations derived above contribute
to the thermodynamics of liquids. Indeed, the free energy can be written
as \begin{equation}
F=F_{0}+F_{1}+F_{2}=F_{0}+T\sum_{\mathbf{q}}\ln\left(1-e^{-\hbar\omega_{1}/T}\right)+T\sum_{\mathbf{q}}\ln\left(1-e^{-\hbar\omega_{2}/T}\right)\,\,\,,\label{40}\end{equation}
where $F_{0}$ is the free energy associated with the particle movements
and $\omega_{1,2}$ are given by equation (\ref{17}). The evaluation
of integrals in equation (\ref{40}) depends on the particular magnitude
of the excitation spectrum, but usually the integrals are rapidly
convergent and their contribution to the thermodynamic properties
of the liquid is small. For instance, the sound-like contribution
is approximately given by $F_{2}\simeq-\pi^{2}VT(T/\hbar v_{s})^{3}/90$,
which is indeed a small correction to $F_{0}$ (the latter being governed
mainly by the liquid cohesion). 

In conclusion, we have shown that in interacting molecular systems
there may appear sound-like excitations controlled by short-range
interactions, distinct from the ordinary hydrodynamic sound. The former
are non-eequilibrium excitations, while the latter appear through
equilibrium, adiabatic processes. The velocity $v_{s}$ of the sound-like
excitations is independent of temperature, while the velocity $v_{0}$
of the ordinary sound depends on temperature, through the adiabatic
compressibility. In order to distinguish them we propose to call the
former \char`\"{}kinetic\char`\"{} modes of particle density, or \char`\"{}densitons\char`\"{}.
In addition, in the presence of Coulomb interaction, the well-known
plasmonic branch is present in the spectrum of the density excitations,
which, for shorter wavelengths and weak Coulomb coupling may look
like another, anomalous, fast sound. We have shown that the \char`\"{}two-sounds
anomaly\char`\"{}  reported in liquids like water, rare-gas mixtures,
metallic alloys, etc, and documented by molecular dynamics studies,
can be understood on this basis.

\end{document}